\documentclass[useAMS,usenatbib,a4paper,fleqn]{mn2e}

\usepackage{amsfonts,amssymb,amsmath}
\usepackage{aas_macros}
\usepackage{times,varioref,multirow,textcomp,comment}
\usepackage[usenames,dvipsnames,svgnames,hyperref]{xcolor}
\usepackage[pdftex]{graphicx}
\usepackage{comment}
\usepackage[
    pdftex,
    a4paper=false,
    plainpages=true,
    pdfpagelabels,
    breaklinks=true,
    bookmarks=true,
    bookmarksopen=false,
    bookmarksopenlevel=2
    bookmarksnumbered=true,
    bookmarkstype=toc,
    colorlinks=true,
    citecolor=RoyalBlue,
    linkcolor=ForestGreen,
    menucolor=Teal,
    urlcolor=DarkOrange,
]{hyperref}

\providecommand{\adsurl}[1]{\href{#1}{ADS}}

\def \Msun{\ {\rm M_\odot}}
\def \Msunh{\ h^{-1}{\rm M_\odot}}

\def \Mpch{\ h^{-1}{\rm Mpc}}

\def \LCDM{$\Lambda$CDM}

\def \rvmax{R_{V_{\rm max}}}
\def \vmax{V_{\rm max}}

\newcommand{\Figref}[1]{Fig.~\ref{#1}}
\newcommand{\Secref}[1]{\S\ref{#1}}  
\newcommand{\Tableref}[1]{Table~\ref{#1}}

\defcitealias{nfw}{NFW}

\defcitealias{elahi2011}{{\sc stf}}

\def \velociraptor{{\sc veloci}raptor}
\defcitealias{ahf}{{\sc ahf}}

\def \Gadget2{{\sc gadget-2}}

\hypersetup{
  pdfauthor = {Pascal Elahi},
  pdfkeywords = {Coupled Dark Sector},
  pdftitle = {Coupled Dark Sector (Sub)Haloes}
}

\def \qCDM{$\phi$CDM}
\newcommand{\dedm}[1]{$\phi(\beta=#1)$CDM}
\def \ncosmo{3}

\begin{document}
\title[Coupled Dark Sector (Sub)Haloes]{Hidden from view: Coupled Dark Sector Physics and Small Scales}
\author[P.J.~Elahi, {\it et al}.]{
\parbox{\textwidth}{
Pascal J. Elahi\thanks{E-mail: pelahi@physics.usyd.edu.au}$^{1}$,
Geraint F. Lewis$^{1}$,
Chris Power$^{2}$,
Edoardo Carlesi$^{3}$,
Alexander Knebe$^{4,5}$,
}\vspace{0.4cm}\\ 
\parbox{\textwidth}{
$^{1}$Sydney Institute for Astronomy, School of Physics, A28, The University of Sydney, NSW 2006, Australia\\
$^{2}$International Centre for Radio Astronomy Research, University of Western Australia, 35 Stirling Highway, Crawley, WA 6009, Australia\\
$^{3}$Racah Institute of Physics, Hebrew University, Jerusalem 91904, Israel\\
$^{4}$Departamento de F\'isica Te\'{o}rica, M\'{o}dulo 15, Facultad de Ciencias, Universidad Aut\'{o}noma de Madrid, 28049 Madrid, Spain\\
$^{5}$Astro-UAM, UAM, Unidad Asociada CSIC
}
}
\maketitle

\pdfbookmark[1]{Abstract}{sec:abstract}
\begin{abstract}
We study cluster mass dark matter haloes, their progenitors and surroundings in an coupled Dark Matter-Dark Energy model and compare it to quintessence and \LCDM\ models with adiabatic zoom simulations. When comparing cosmologies with different expansions histories, growth functions \& power spectra, care must be taken to identify unambiguous signatures of alternative cosmologies. Shared cosmological parameters, such as $\sigma_8$, need not be the same for optimal fits to observational data. We choose to set our parameters to \LCDM\ $z=0$ values. We find that in coupled models, where DM decays into DE, haloes appear remarkably similar to \LCDM\ haloes despite DM experiencing an additional frictional force. Density profiles are not systematically different and the subhalo populations have similar mass, spin, and spatial distributions, although (sub)haloes are less concentrated on average in coupled cosmologies. However, given the scatter in related observables ($\vmax,\rvmax$), this difference is unlikely to distinguish between coupled and uncoupled DM. Observations of satellites of MW and M31 indicate a significant subpopulation reside in a plane.  Coupled models do produce planar arrangements of satellites of higher statistical significance than \LCDM\ models, however, in all models these planes are dynamically unstable. In general, the nonlinear dynamics within and near large haloes masks the effects of a coupled dark sector. The sole environmental signature we find is that small haloes residing in the outskirts are more deficient in baryons than their \LCDM\ counterparts. The lack of a pronounced signal for a coupled dark sector strongly suggests that such a phenomena would be effectively hidden from view. 
\end{abstract}
\maketitle

\section{Introduction}\label{sec:intro}
The \LCDM\ concordance model of cosmology describes a spatially flat universe with an energy budget dominated by a invisible Dark Sector composed of two major components: Dark Matter (DM) that governs the small scale clustering of luminous matter; and Dark Energy (DE), which drives the late time accelerated expansion. This model is based on multiple lines of evidence, from the Cosmic Microwave Background (CMB) anisotropies \cite[e.g.][]{wmap9,planckcosmoparams2013,planckcosmoparams2015}, Baryonic Acoustic Oscillations (BAO) \cite[e.g]{beutler2011a,blake2011a,anderson2014a}, Large-Scale Structure (LSS) \cite[e.g.][]{abazajian2009a,beutler2012a}, weak lensing \cite[e.g.][]{kilbinger2013a,heymans2013a}, cluster abundances \cite[e.g][]{vikhlinin2009a,rozo2010a}, galaxy clustering \cite[e.g.][]{tegmark2004,reid2010a}, to the luminosity-distance relation from Type Ia \cite[e.g.][]{kowalski2008,conley2011a,suzuki2012a}. Despite all this evidence, the nature of the Dark Sector remains one of the most fundamental mysteries in cosmology. Observational evidence strongly favours nonbaryonic elementary particle DM, though the exact properties of the DM particle(s) are poorly constrained. Most studies have focused on so-called Cold Dark Matter (CDM) \cite[see][for a review]{frenkwhite2012}, which has well-motivated candidates from particle physics, {\it e.g.} the lightest supersymmetric particle or neutralino (e.g.~\citealp{ellis1984}; or for a summary of several candidates see \citealp{bertone2005,petraki2013}). Similarly, the measured expansion rate at late times implies that the DE is well characterised by a cosmological constant, $\Lambda$, with an measured equation-of-state (eos) $w\equiv\rho/p\approx-1$, although the fact that the Universe is not exactly homogeneous may introduce bias in these observations \cite[e.g.][]{clarkson2012}.

\par
In spite of the remarkable agreement of \LCDM\ with observations {\em of large-scale structure}, the model is in tension with observations on galaxy scales, the most well known of which is the so-called ``missing satellite problem'': CDM models predict many more satellite galaxies than observed around galaxies such as our own \cite[e.g.][]{klypin1999,moore1999}. The excess subhaloes that do not host galaxies may indicate that feedback processes, such as surpernovae, efficiently remove gas from low mass subhaloes leaving them almost completely dark \cite[e.g.][]{bullock2000,benson2002,nickerson2011,nickerson2012}, or may indicate the need for modifications to CDM, such as Warm Dark Matter \cite[e.g.][]{lovell2013,schneider2013b,power2013a,elahi2014a}. Perhaps the most intriguing observational conundrum is the as yet unexplained alignments of satellite galaxies surrounding the Milky Way and its galactic neighbour, Andromeda, the so-called Vast Polar Structure \cite[VPOS][]{pawlowski2012} and Plane of Satellites \cite[][]{ibata2013,conn2013} respectively. This alignment is not unique to the local group; alignments have observed in SDSS data \cite[e.g.][]{yang2006,li2013a}, and observations show satellite galaxies have velocities that are significantly more correlated than theory would predict (e.g.,\citealp{ibata2014b,gillet2015}; for counterarguments see \citealp{sawala2015a,libeskind2015a}).

\par
There are also some theoretical shortcomings with \LCDM, such as the fine-tuning and coincidence problems, which remain poorly explained from a purely theoretical point-of-view \cite[see for instance][]{ferreira1997,brax2000,huey2006}. The former problem refers to the fact that, if we assume that the DE is a cosmological constant arising from the zero-point energy of a fundamental quantum field, then its density requires an unnatural fine-tuning of several tens of orders of magnitude to be compatible with observed cosmological constraints. The latter problem arises from the difficulty in satisfactorily explaining the observation that matter and DE energy densities today have comparable values given that the energy density of these two fluids have completely different dependencies on cosmic time and only now is DE beginning to dominate. If DE dominated at early times, cosmic structure formation would be strongly suppressed. 

\par
This coincidence problem has lead to the proposal of several alternatives, such as modifications to general relativity (e.g. \citealp{hu2007,starobinsky1980}; and \citealp{defelice2010} for a review) or dynamical scalar fields (e.g. \citealp{ratra1988,wetterich1988,armendarizpicon2001}; and \citealp{tsujikawa2013a} for a review). Dynamical scalar fields, or quintessence models, are an attractive alternative since, in principle, they can alleviate some of the fine-tuning problem. An interesting subset of these models are those where the scalar field couples to the matter sector, thereby removing the coincidence problem \cite[e.g.][]{wetterich1995,amendola2000}. Typically it is assumed that the dark energy couples only to dark matter, be it cold or warm, (for lack of evidence indicating normal matter interacting with a hidden sector). The coincidence problem is alleviated by having the dark matter decay to the scalar field producing the late time accelerated expansion. 

\par
In N-Body realisations of these models, this interaction gives rise to two novel effects: the DM N-Body particle masses vary with time; and the effective gravity constant governing two-body interactions between DM N-Body particles is no longer the same as that governing baryon-DM or baryon-baryon interactions. Studies have examined cosmological structure formation using N-Body simulations for both uncoupled \cite[e.g.][]{klypin2003a,dolag2004a,deboni2011a} and coupled models \cite[e.g.][]{maccio2004a,gadgetcodecs,baldi2010,li2011a}. \cite{li2011a} showed that there are significant differences in the matter power spectrum as a result of the different growth functions. \cite{sutter2015a} found void density profiles and the number of very large voids are affected by coupled cosmologies, although this study compared \LCDM\ to a very strongly coupled model, one that is ruled out by observations \cite[][]{pettorino2012} and the differences are small. \cite{pace2015} \& \cite{giocoli2015a} found differences in the weak lensing signature of coupled models. However, \cite{carlesi2014a,carlesi2014b} found negligible differences in the cosmic web and halo mass function between \LCDM\, uncoupled and coupled models. However, these authors did find weak differences in the concentration and spin parameter of small (moderately resolved) field haloes.

\par
The hint that (un)coupled quintessence models statistically differ from \LCDM\ at not only large scales but for small haloes is exciting given the tensions that exist between observations and \LCDM\ predictions. Our aim is to examine the distribution of dark matter haloes in hydrodynamical zoom simulations, identify the observational fingerprints of a coupled dark sector and explore whether these models can reduce the observational tensions that exist with the current concordance \LCDM\ model. This paper is organised as follows: we describe the numerical methods in \Secref{sec:methods} and discuss some issues concerning comparing different cosmological models. Our findings are presented in sections \ref{sec:halo}-\ref{sec:fieldhaloes}. These results are discussed in \Secref{sec:discussion}.

\section{Methods}\label{sec:methods}
\subsection{The model}
As the theoretical and numerical basis for coupled quintessence models has been thoroughly covered before we only briefly describe the cosmological model here (for theoretical and observational discussions see \citealp{amendola2000,tsujikawa2013a}; for discussions of numerical approaches see for instance \citealp{maccio2004a,gadgetcodecs,li2011a,carlesi2014a}). Dark energy in these models arises from the evolution of a scalar field, $\phi$, whose Lagrangian is generically written as:
\begin{align}
  L=\int d^4x \sqrt{-g}\left[-\tfrac{1}{2}\partial_\mu\partial^\mu\phi+V(\phi)+m(\phi)\psi_m\bar{\psi}_m\right],
\end{align}
that is a kinetic term, a potential term and a term characterising the coupling of the scalar field to the matter sector. There are several models for the potential, $V(\phi)$ that give rise to late time accelerated expansion, mimicking the affect of a cosmological constant $\Lambda$. Here we focus on the so-called Ratra-Peebles potential \cite[][]{ratra1988}:
\begin{align}
  V(\phi)=V_o\phi^{-\alpha},
\end{align}
where $\phi$ is in units of the Planck mass. 

\par
There is a great deal of freedom in choosing the form of $m(\phi)$, the interaction term. A popular, simple model that can alleviate the coincidence problem is an exponential coupling of the scalar field to {\em dark matter only}
\begin{align}
  m(\phi)=m_o\exp\left[-\beta(\phi)\phi\right].
\end{align}
For simplicity, we use a constant coupling term, $\beta(\phi)=\beta_o$. The first consequence of this coupling is the decay of dark matter particles into the scalar field. A second consequence is that dark matter and baryonic matter are governed by different dynamics. Whereas baryons follow standard Newtonian dynamics, dark matter experiences an additional frictional force which appears as an effective gravitational force, $G_{\rm eff}=G(1+\beta_o^2)$. The two matter fluids develop an offset in the amplitude of their density perturbations, a 'gravitational bias' that can significantly impact the baryon fraction of galactic- or cluster-sized haloes \cite[e.g.][]{baldi2010}. 

\subsubsection{Initial conditions and comparing cosmologies}
Before discussing the simulations themselves it is worthwhile considering the initial conditions and how cosmological parameters such as the matter density are set. Different cosmological observations tightly constrain different sets of cosmological parameters, with each observation having different correlations between various parameters. CMB measurements, for instance, primarily constrain quantities such as the amplitude of the matter power spectrum at a moment in time during the matter dominated era when the last scattering surface is produced \footnote{This is not strictly true as distortions along the line-of-sight, from weak lensing for example, affects the temperature power spectrum of the CMB}. Supernovae observations measure the expansion rate over cosmic time at (moderately) low redshifts. Moreover, observational data is only meaningful in the context of a model. As a consequence, some models and their associated parameters are not necessarily tightly constrained if one only considers a single observational data set.

\par 
An alternative model may share cosmological parameters with \LCDM, such as $\sigma_8(z=0)$, the normalisation of the mass variance quantifying the nonlinearity of regions enclosing a fixed amount of mass. However, analysing the data with a particular model will lead to different optimal values for these shared parameters, which may not even agree to within some confidence level. This gives some flexibility in setting the cosmological parameters when running simulations of alternative models. One could choose to set the values of the shared subset based on estimates derived from the CMB interpreted through the \LCDM\ lens. A consequence of this choice is that the parameters could have far different $z=0$ values in the alternative model when compared to \LCDM. An example of the above are the coupled models studied by \cite{gadgetcodecs}. This author fixed the power spectrum amplitude at $z_{\rm CMB}$, $\mathcal{A}_{\rm s}$, across all the models but $\Omega_{\rm b}$ \& $\Omega_{\rm m}$ are set to be equal at $z=0$. The result is that $\sigma_8$ differs significantly between models, from $0.8-0.9$. In contrast, the matter density parameters start at different values in order to agree at $z=0$ with a \LCDM\ interpretation of CMB observations. 

\par
Here we have taken a slightly different approach, used in \cite{carlesi2014a}, and set both $\sigma_8(z=0)$ and the matter density parameters based on CMB observations by Planck interpreted with a \LCDM\ model \cite[][]{planckcosmoparams2013}. As a consequence, the amplitude of the density perturbations matches at $z=0$ but starts from a different point at $z_{\rm CMB}$ and $z_{\rm i}$, the redshift were we start our simulations. Neither approach is more correct than the other. However, it is important to note that differences will arise between cosmological models due to ``simple'' differences in cosmological parameters like $\sigma_8$ and $\Omega_m$ that may hide differences arising from alternative physics or differences in the growth of density perturbations \footnote{See \cite{pace2015}, who show the change in weak lensing signals from a \LCDM\ model with the same $\sigma_8(z=0)$ as an alternative model accounts for a significant portion of the signal in the alternative model.} We note that although methods exist that map the results from a simulated cosmology to another, producing the same halo mass function, the resulting mapped haloes have systematically biased internal properties, i.e., the nonlinear evolution is not correctly accounted for \cite[e.g.][]{angulo2010a,mead2014a}. Thus we do not attempt to estimate the change in a given property is a result of different cosmological parameters, rather we attempt to identify properties that appear affected by the inclusion of extra dark sector physics.

\subsection{Simulations}
\begin{figure*}
    \centering
    \includegraphics[width=0.96\textwidth, trim=2.0cm 15.0cm 2.0cm 3.0cm, clip=true]{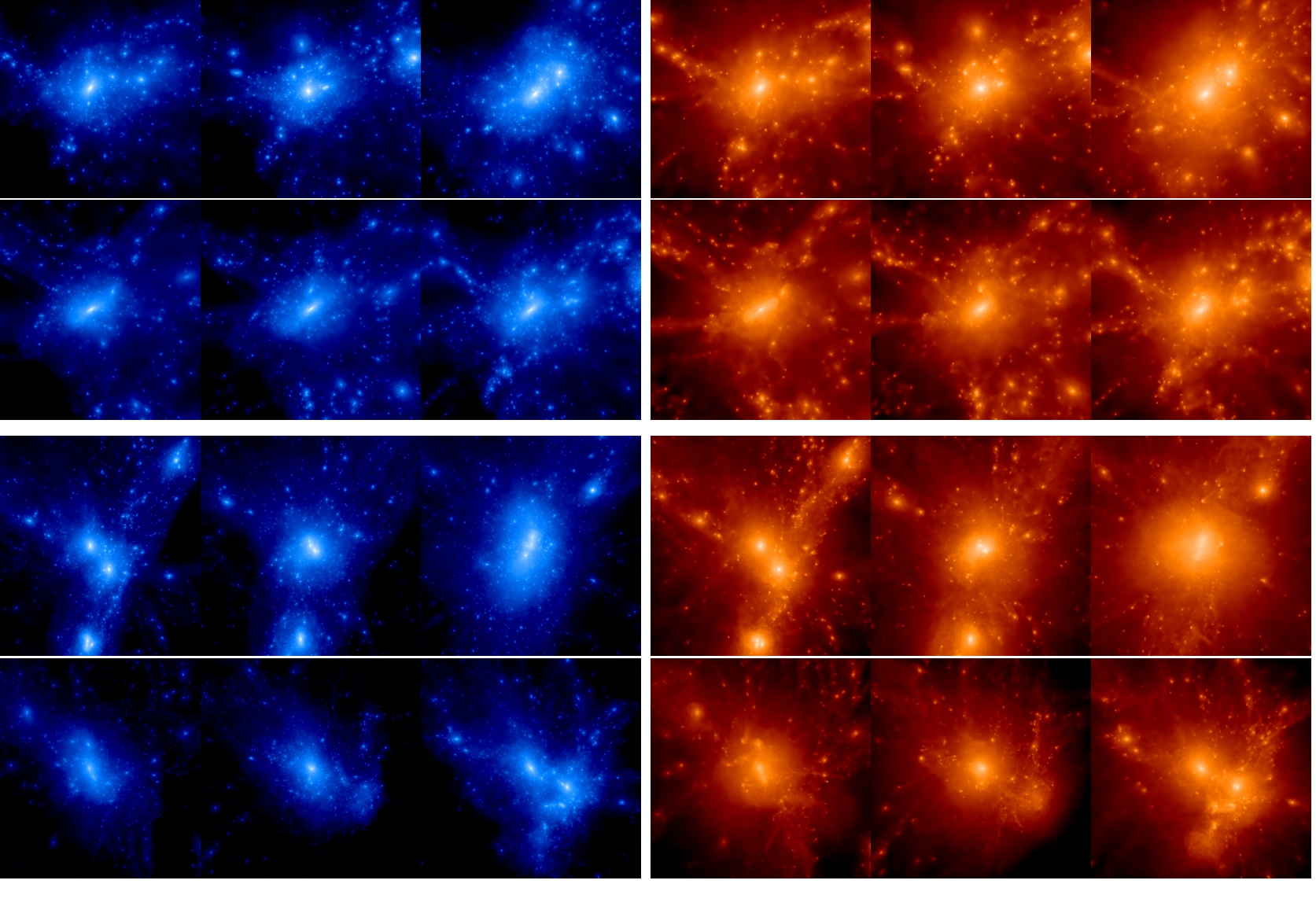}
    \caption{Smoothed projected density images of the high resolution region at $z=0$ centred the two groups used in this study. Q (top two rows) \& MM (bottom two rows). From left to right: \LCDM, \qCDM, \dedm{0.05}. Top row in each subpanel shows CDM distribution and bottom row shows gas distribution. Densities are plotted on a logarithmic scale with bright regions being dense. The scale is arbitrary but kept fixed for a given particle type and halo.}
    \label{fig:sims}
\end{figure*}

We use \ncosmo\ cosmologies in this study: a reference \LCDM; an uncoupled quintessence model (\qCDM); and one coupled model ($\beta_o=0.05$). As mentioned previously all cosmologies have parameters consistent with $z=0$ \LCDM\ Planck data $(h,\Omega_m,\Omega_b,\sigma_8)=(0.67,0.3175,0.049,0.83)$ \cite[][]{planckcosmoparams2013,planckcosmoparams2015}. Our coupled model is chosen to test the boundaries of allowed coupling \cite[see][]{pettorino2012} in order to maximise any observational differences that may exist between this cosmology and the standard \LCDM\ model. The power spectrum and evolution of these quantities along with the growth factor $f\equiv d\ln D(a)/d\ln a$ are calculated using first-order Newtonian perturbation equations and the publicly available Boltzmann code {\sc cmbeasy} \cite[][]{cmbeasy}. Initial conditions are produced using a uniform Cartesian grid and a first-order Zel'Dovich approximation using a modified version of the publicly available {\sc n-genic} code. The modified code uses the growth factors and expansion history calculated by {\sc cmbeasy} to correctly calculate the particle displacements in the non-standard cosmologies. 

\par
We produce several well resolved dark matter haloes using hydrodynamical zoom simulations in the cosmologies mentioned previously. As the N-Body code used here is described in detail in \cite{carlesi2014a}, we briefly summarise the key points here. {\sc dark-gadget} is a modified version of {\sc p-gadget}-2. The key modifications are the inclusion of a separate gravity tree to account for the additional long range forces arising from the scalar field, and an evolving dark matter N-body particle mass which models the decay of the dark matter density. The code requires the full evolution of the scalar field, $\phi$, the mass of the dark matter N-body particle, and the expansion history. 

\par
The zoom simulations used a parent simulation of $L_{\rm box}=50\Mpch$ containing $2\times256^3$ particles (DM and gas particles) with the following cosmological parameters. Candidate objects were identified using \velociraptor\ \cite[][]{elahi2011}. For each selected object, all particles within a radius of $\sim 3 R_{\rm max}$ of the halo at $z=0$ are identified in the high resolution IC that was down sampled to produce the low resolution parent simulation. The mass resolution in the zoom region is $8\times10^{6}\Msunh$ and $1.5\times10^6\Msunh$ for dark matter\footnote{This is the dark matter particle mass at $z=0$ for the coupled dark sector simulations.} and gas particles respectively, or an effective resolution of $1024^3$. All the simulations are started at $z=65$ with the same phases in the density perturbations and use a gravitational softening length of $1/40$ of the interparticle spacing.  We study our haloes across cosmic time, focusing specifically on $z=0$ and $z=0.5$, where the progenitors of the clusters are large galaxy groups. The cosmological parameters relative to the fiducial \LCDM\ model at $z=0.5$, where the quantities are not normalised, are listed in \Tableref{tab:cosmoparams}.
\begin{table}
\setlength\tabcolsep{2pt}
\centering\footnotesize
\caption{Comparison of several key cosmological parameters relative \LCDM\ at $z=0.5$. Here $m_{\rm DM}$ is dark matter mass, $D_+$ is the growth function, $H$ is the Hubble expansion rate, and $G_{\rm eff}$ is the effective gravitational potential in DM-DM interactions.}
\begin{tabular}{l|ccccc}
\hline
\hline
    Cosmology &
    $\frac{m_{\rm DM}}{m_{{\rm DM}}^{\Lambda{\rm CDM}}}$ & $\frac{\Omega_{\rm cdm}}{\Omega_{{\rm CDM}}^{\Lambda{\rm CDM}}}$ & $\frac{D_+}{D_{+}^{\Lambda{\rm CDM}}}$ & $\frac{H}{H^{\Lambda{\rm CDM}}}$ & $\frac{G_{\rm eff}}{G_{{\rm eff}}^{\Lambda{\rm CDM}}}$ \\
\hline
    \qCDM           & 1         & 0.979 & 0.992 & 1.0129 & 1        \\
    \dedm{0.05}     & 1.0027    & 0.993 & 1.009 & 1.0024 & 1.0025   \\
\end{tabular}
\label{tab:cosmoparams}
\end{table}

\par
We identify all bound haloes and their substructures using \velociraptor\ within the high resolution region. This code identifies field haloes using a 3D Friends-of-Friends (FOF) algorithm with a linking length of $0.2$ times the inter-particle spacing\footnote{We also apply a 6DFOF to each candidate FOF halo using the velocity dispersion of the candidate object to clean the halo catalogue of objects spuriously linked by artificial particle bridges, useful for disentangling mergers}. Substructures are then found by identifying particles that belong to a local velocity distribution that differs from average background and then linking this outlier population using a phase-space FOF approach \cite[for details see][]{elahi2011,elahi2013a}. This technique not only identifies bound subhaloes, it will also identify the tidal features associated with the subhalo and even completely unbound tidal debris. 

\section{Haloes in Coupled Cosmologies}\label{sec:halo}\
\begin{table*}
\setlength\tabcolsep{2pt}
\centering\footnotesize
\caption{Bulk Properties of Q \& MM: total bound mass $M_{\rm tot}$; total number of particles $N_p$; mass within the so-called virial radius $M_{\Delta}=4\pi\Delta\rho_{\rm bg}R_\Delta^3/3$, where $\Delta=200$; baryon fraction $f_{\rm b}$; maximum circular velocity $\vmax$; radius of maximum circular velocity $\rvmax$; number of subhaloes within the virial radius $N_{\rm subs}$; mass fraction in subhaloes $f_{M, {\rm subs}}$; total number of substructures (i.e., both intact and tidally disrupted subhaloes) within the virial radius $N_{\rm S}$; mass fraction in substructures $f_{M,{\rm S}}$; mass fraction in unbound tidal debris $\Delta f_{M,{\rm S,tid}}$.}
\begin{tabular}{l|l| cccccccc ccccc}
\hline
\hline
     & Cosmology 
     & $M_{\rm tot}$ & $N_{\rm p, DM}$ & $M_{\Delta}$\footnotemark& $f_{\rm b}$ & $\vmax$ & $\rvmax$ & $R_{\rm \Delta}$ & $c_{\vmax}$ & $N_{\rm subs}$ & $f_{M,{\rm subs}}$ & $N_{\rm S}$ & $f_{M,{\rm S}}$ & $\Delta f_{M,{\rm S,tid}}$\\
     & & [$10^{13}\Msun$] & $10^6$ & [$10^{13}\Msun$] & & km/s & kpc & kpc & & & \\
\hline
    \multirow{4}{*}{Q($z=0$)}
    & \LCDM          & 8.25  & 5.41  & 6.44  & 0.132 & 687   & 177 & 865  & 4.89 & 1036  & 0.069 & 1050 & 0.098 & 0.029\\
    & \qCDM          & 9.39  & 6.18  & 7.97  & 0.125 & 743   & 265 & 930  & 3.51 & 1454  & 0.097 & 1485 & 0.135 & 0.038\\
    & \dedm{0.05}    & 16.00 & 10.73 & 11.91 & 0.119 & 826   & 314 & 1063 & 3.38 & 2544  & 0.122 & 2599 & 0.198 & 0.078\\
\hline
    \multirow{4}{*}{Q($z=0.5$)} 
    & \LCDM          & 4.29  & 2.85  & 3.40  & 0.139 & 629   & 118 & 470 & 3.96 & 633  & 0.047 & 658  & 0.079 & 0.032\\
    & \qCDM          & 4.20  & 2.76  & 3.80  & 0.126 & 673   & 193 & 486 & 2.51 & 529  & 0.071 & 542  & 0.102 & 0.031\\
    & \dedm{0.05}    & 7.52  & 4.97  & 6.04  & 0.130 & 778   & 248 & 567 & 2.28 & 960  & 0.080 & 991  & 0.119 & 0.039\\
\hline\hline
    \multirow{4}{*}{MM($z=0$)} 
    & \LCDM          & 3.11  & 2.09  & 2.94  & 0.126 & 518 & 207 & 667  & 3.21 & 445  & 0.033 & 467   & 0.094 & 0.059\\
    & \qCDM          & 7.04  & 4.63  & 6.39  & 0.127 & 668 & 393 & 864  & 2.19 & 1434 & 0.128 & 1450  & 0.244 & 0.116\\
    & \dedm{0.05}    & 13.85 & 9.43  & 11.70 & 0.117 & 708 & 774 & 1057 & 1.36 & 2293 & 0.077 & 2336  & 0.113 & 0.036\\
\hline
    \multirow{4}{*}{MM($z=0.5$)}
    & \LCDM          & 2.42  & 1.62  & 1.94  & 0.115 & 501 & 211 & 381 & 1.80 & 335   & 0.118 & 356   & 0.177 & 0.059\\
    & \qCDM          & 2.40  & 1.61  & 2.20  & 0.123 & 553 & 170 & 398 & 2.33 & 315   & 0.068 & 335   & 0.142 & 0.074\\
    & \dedm{0.05}    & 6.07  & 4.07  & 3.10  & 0.117 & 621 & 165 & 446 & 2.70 & 385   & 0.061 & 415   & 0.102 & 0.039\\
\end{tabular}
\label{tab:bulkproperties}
\end{table*}
\footnotetext{We use the $\Delta\rho_{\rm bg}=\Delta\Omega_{m,o}\left(3H^2(a)/8\pi G\right)$ \LCDM\ definition, that is here we are not explicitly taking into account the differences in the expansion history when determining the overdensity that constitutes a virialised system in the non-standard cosmologies.} 

Here we focus on two objects: one with a quiescent history (Q); and one which is undergoing and has undergone several major mergers (MM). These two objects are chosen to explore the effects of coupled dark sector physics in two different dynamical regimes. We show the dark matter \& gas distributions at $z=0.0$ \& $z=0.5$  in \Figref{fig:sims} and list their bulk properties in \Tableref{tab:bulkproperties}. 

\par 
Focusing on \Figref{fig:sims}, we see that the quiescent group looks remarkably similar at all times in spite of the fact that \dedm{0.05}-Q halo has a maximum circular velocity that $20\%$ higher and is $\approx80\%$ more massive at $z=0$ that its counterparts in uncoupled simulations. Considering the how self-similar dark matter haloes appear, this fact in itself is not very surprising. 

\par 
MM on the other hand appears to be at different stages of a major merger. In \dedm{0.05}, at $z=0$, the halo is relaxing from the major merger event that occurred at $z=0.5$. Three ``cores'' are visible, remnants of past mergers. The ongoing merger at $z=0.5$ is the reason for the large differences in the total bound FOF mass of the halo relative to its virial mass seen in \Tableref{tab:bulkproperties}. The MM halo in uncoupled cosmology also has several cores at $z=0$ and is near a similar mass halo which which it will merge in the near future. On the other hand, \LCDM-MM has yet to undergo its first major merger, as evidence by its lack of merger remnants. 

\par 
Clearly for our choice of cosmological parameters, haloes at these mass scales are more massive in strongly coupled cosmologies but can have similar masses in cosmologies with the same dark matter density and physics but different expansion histories and power spectra. Hence, the increase in mass is a result of higher matter densities (particularly at early times), a higher effective gravitational force experienced by dark matter and higher growth functions (see \Tableref{tab:cosmoparams}). This difference is in spite of the fact that these parameters change by a few percent at most, and $\sigma_8(z=0)$ is the same in every cosmology. However, larger haloes is not an unambiguous signal of a coupled dark sector. Despite the differences in mass {\em at this particular scale and time}, the freedom allowed in the $(\Omega_m,\sigma_8)$ plane means that the current differences in the mass of these objects is not necessarily a useful signature of coupled dark matter. For both Q \& MM, moving from one cosmology to the next appears to look like viewing the same object at different times. Therefore, it is quite possible that a different viable choice of $(\Omega_m,\sigma_8)$ would remove these differences. 

\par 
None of the properties listed in \Tableref{tab:bulkproperties} show an unambiguous trend with coupling that is not affected by the merger history. For example, the concentration parameter was found by \cite{carlesi2014a} to be lower in coupled cosmology. This trend is followed by Q but not by MM at early times. Moreover, even the uncoupled cosmology has a lower concentration for these haloes. The fraction of the host's mass bound up in subhaloes also varies between cosmologies as does the fraction in tidal debris. 

\par 
Even sampling the growth of structure {\rm across some fraction of cosmic time} need not be necessarily effective for the same reason that as the observed mass difference. The weak gravitational lensing study by \cite{pace2015} \& \cite{giocoli2015a} found differences in the convergence power spectrum between \LCDM\ and coupled cosmologies, however, much of the difference arises to different $\sigma_8$ normalisations. We therefore study the internal properties of these objects in search of unambiguous probes of coupled dark sector physics. 

\subsection{Profiles}
\begin{figure}
    \centering
    \includegraphics[width=0.45\textwidth, trim=1.75cm 15.2cm 11.5cm 2.5cm, clip=true]{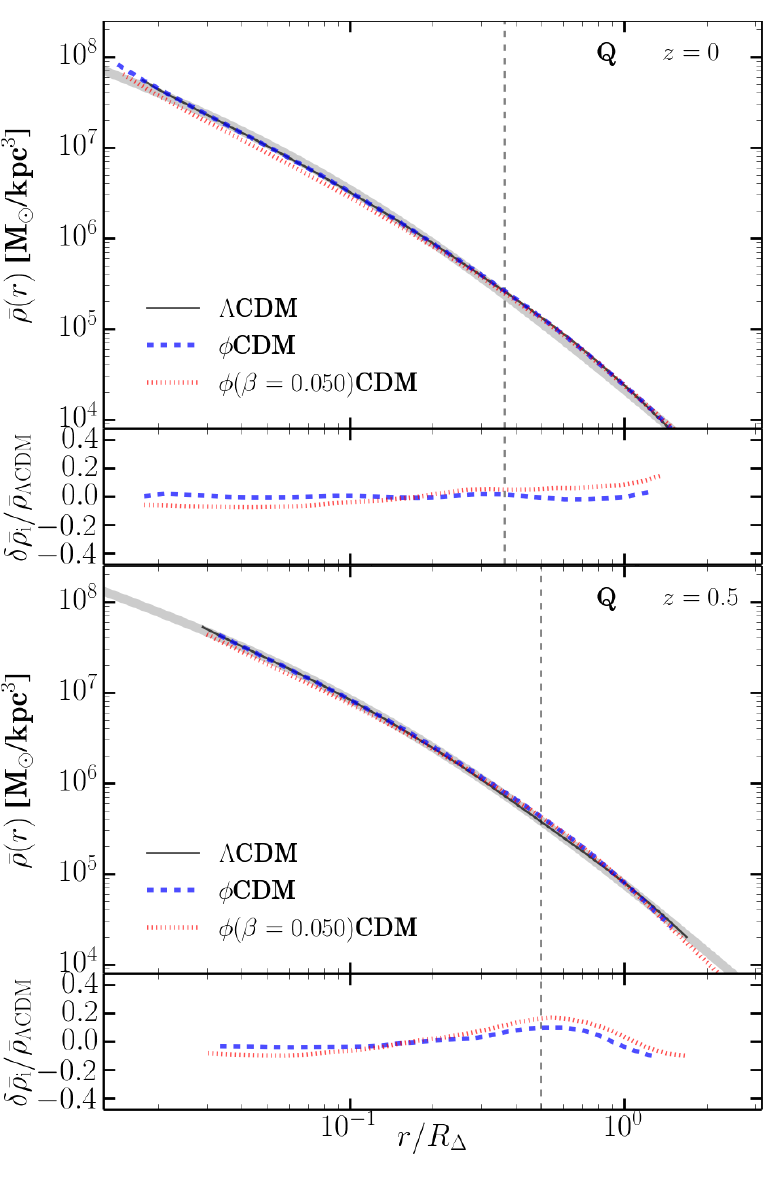}
    \caption{Radial density profiles along with residuals (where we have removed the average offset to emphasise changes in slope) of Q at $z=0$ (top two panels) and at $z=0.5$ (bottom two panels). We show a thick dark gray region showing the best fit NFW profile to the \LCDM\ model and a dashed vertical line at the fitted scale radius.}
    \label{fig:totdenprof:Q}
\end{figure}

The first question is whether the density profiles differ. \cite{carlesi2014b} stacked clusters to measure the average radial density profile at $z=0$ and found negligible differences save in the poorly resolved, inner region for the most strongly coupled cosmology. Based on \Figref{fig:sims}, it is not obvious major differences will exist for the quiescent Q group but there should be differences in the merging MM group, particularly for \dedm{0.05} arising from the dynamical state. The Q quiescent halo profiles shown in \Figref{fig:totdenprof:Q} are almost indistinguishable. The residuals do show that the central density of the strongly coupled cosmology is lower than the \LCDM\ one but key is that the residuals are flat. The shape is not systematically different, all are reasonably characterised by NFW (or Einasto) profile \cite[e.g][]{nfw,navarro2004}.

\begin{figure}
    \centering
    \includegraphics[width=0.45\textwidth, trim=1.75cm 15.2cm 11.5cm 2.5cm, clip=true]{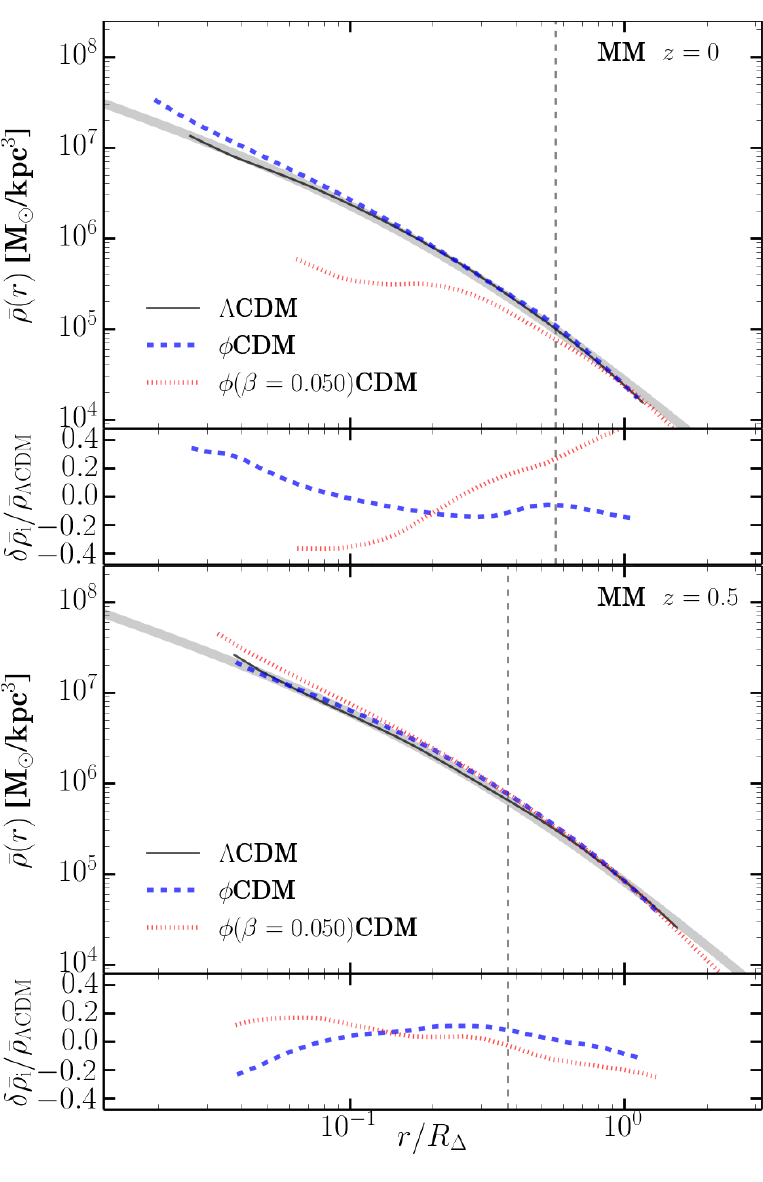}
    \caption{Same as \Figref{fig:totdenprof:Q} but for MM.}
    \label{fig:totdenprof:MM}
\end{figure}

For MM, shown in \Figref{fig:totdenprof:MM}, the effects of the merger are clearly visible. The deviation away from an NFW in the central regions of \dedm{0.05}-MM at $z=0$ are due to the presence of multiple cores (see \Figref{fig:sims}, where the cores for MM are very evident)\footnote{Note the \dedm{0.05}-MM profile in \Figref{fig:totdenprof:MM} is truncated at small radii as a result of 1) the CM, which is calculated by an iteratively using ever smaller radial apertures, lies between two cores and 2) bins must contain $1\%$ of the total number of particles in the halo. The CM lies off a density peak and as a result the first bin extend to larger radii than in the other MM halos. We do not attempt to address this minor issue in this figure as it emphasises the unrelaxed state of the halo.}. A deviation is also seen in the uncoupled model. The profile of \dedm{0.05}-MM at $z=0.5$ also extends to much larger radii, a result of the major merger taking place, with the smaller group residing just past $R_\Delta$. Of greater importance is the similarity in the shape of the profiles at $z=0.5$, when the main halo is relatively undisturbed in all three cosmologies. Again, the cosmologies have produced haloes which are remarkably similar. Even the shallower interior slopes in coupled cosmologies noted for Q and in previous studies of relaxed haloes is not seen here as it has been affected by the dynamical state of the system.

\par
\begin{figure}
    \centering
    \includegraphics[width=0.45\textwidth, trim=1.75cm 14.5cm 11.5cm 2.5cm, clip=true]{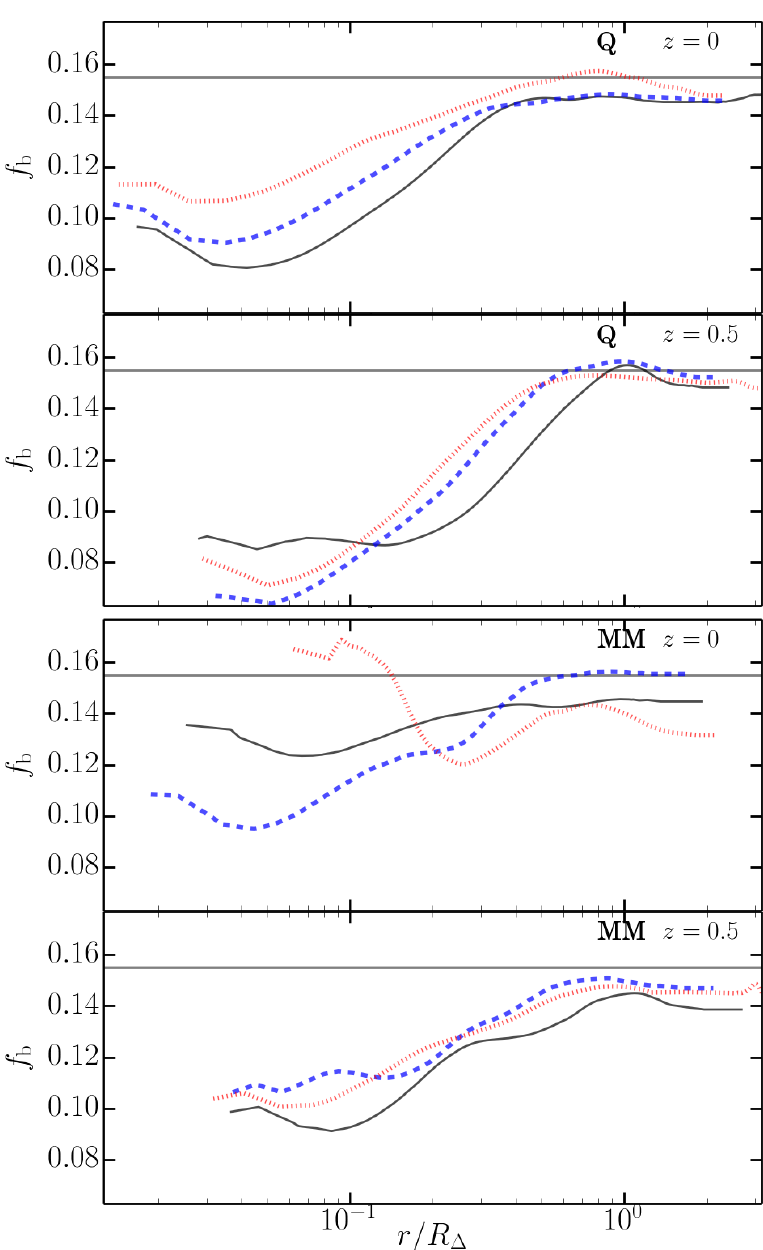}
    \caption{Radial baryon fractions of Q (top two panels, with upper for $z=0$, lower for $z=0.5$) and MM (bottom two). Line styles and colours are the same as in \Figref{fig:totdenprof:Q}.}
    \label{fig:gasprof}
\end{figure}

Given that the effective gravitational field experienced by dark matter differs from that of baryons in coupled cosmologies, one might expect a difference in the distribution of baryons, which is shown in \Figref{fig:gasprof}. Unfortunately, there is a great deal of disparity between the cosmologies and little indication that on an individual object basis, the baryons can be used to discriminate between uncoupled and coupled dark matter physics, even in simplistic adiabatic simulations. By looking at various times and dynamical state, we can clearly see that differences in accretion history are of greater importance than a coupled dark sector. 

\section{Subhaloes: Coupled cosmologies in high density environments}\label{sec:subhaloes}
The bulk host haloes properties show few unambiguous differences between uncoupled and coupled cosmologies. Here we examine if subhaloes, which will host galaxies of similar mass to the Milky Way, show major differences. We first focus on the subhalo population residing within the high density environment inside the virial radius of the host haloes, Q \& MM. 
\subsection{Mass distribution}
\begin{figure}
    \centering
    \includegraphics[width=0.45\textwidth, trim=1.75cm 15.2cm 11.5cm 1.9cm, clip=true]{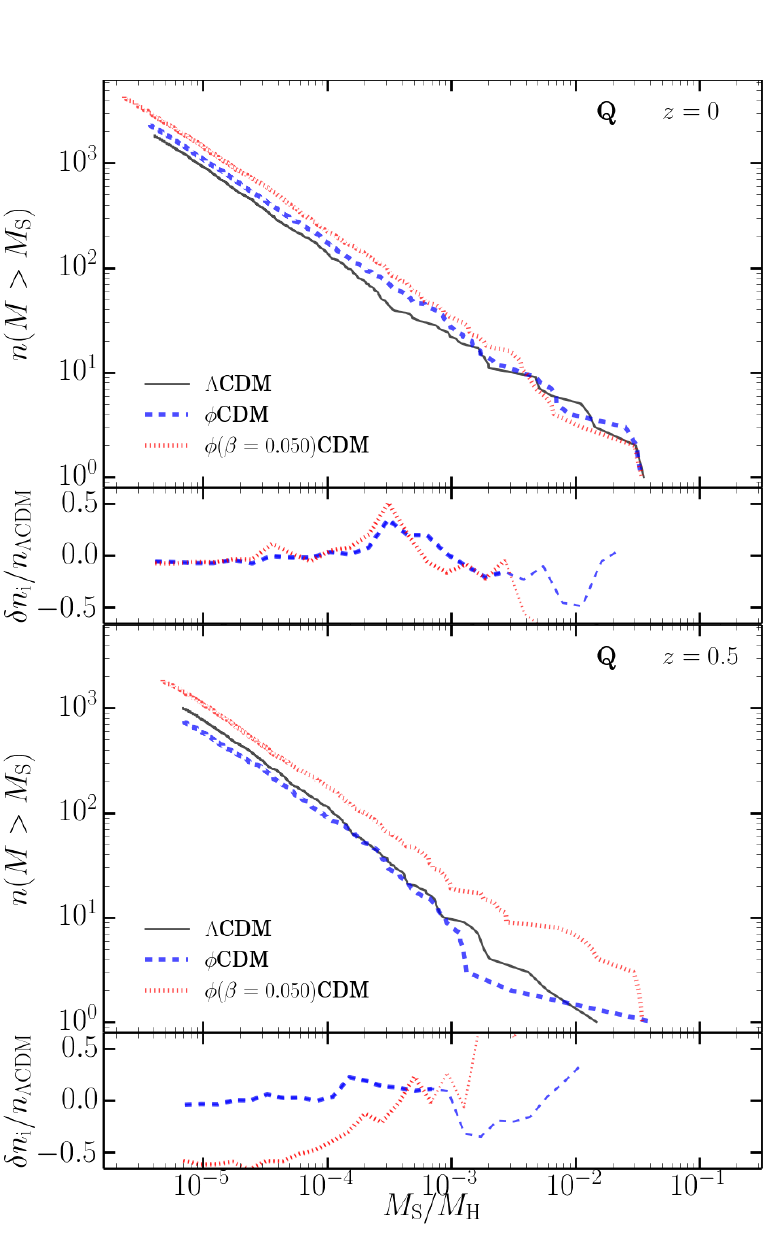}
    \caption{The $z=0$ cumulative mass function along (large panels) with the residuals in the cumulative number (smaller panels) for Q (top two panels) and MM (bottom two panels). In the smaller panels showing the ratio, thin lines correspond to where $n_i$ \& $n_{\Lambda{\rm CDM}}$ $\leq10$, i.e., where the statistics are poor. Line styles and colours are the same as in \Figref{fig:totdenprof:Q}}
    \label{fig:submassfunc:Q}
\end{figure}

\begin{figure}
    \centering
    \includegraphics[width=0.45\textwidth, trim=1.75cm 15.2cm 11.5cm 2.5cm, clip=true]{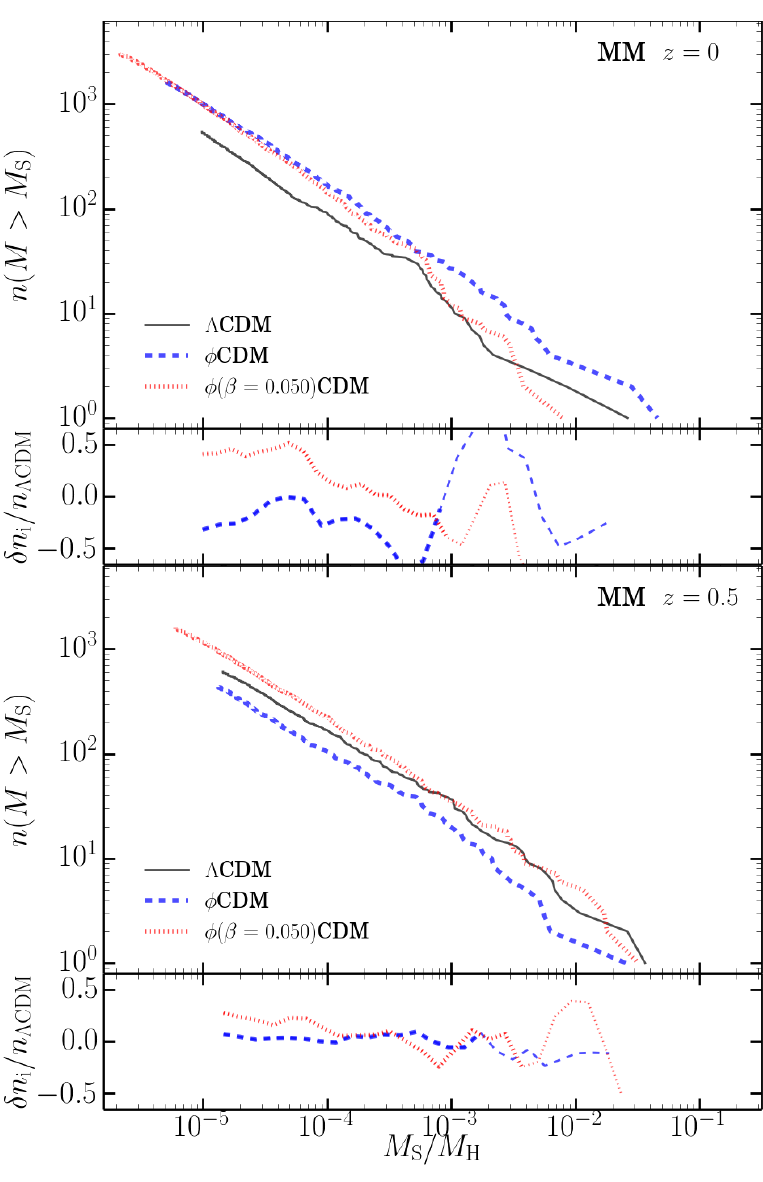}
    \caption{Same as \Figref{fig:submassfunc:Q} but for MM.}
    \label{fig:submassfunc:MM}
\end{figure}

The simplest comparison to make is with the subhalo mass distribution, shown in \Figref{fig:submassfunc:Q}-\ref{fig:submassfunc:MM}. Note that we do {\em not} classify the main merging halo in the \dedm{0.05}-MM simulation at $z=0.5$ as a ``subhalo'' and so it is not present in this figure. Remarkably, the shape of the subhalo mass function is unaffected by the differences in the initial power spectra, expansion histories and, at first glance, even the dynamical state of the host halo! This result is seen in the lower subpanels of each figure. The residual $\delta n_{\rm i}/n_{\rm \Lambda CDM}=n_{\rm i}/n_{\rm \Lambda CDM}-\left\langle n_{\rm i}/n_{\rm \Lambda CDM}\right\rangle$, has the average amplitude difference between models, $\left\langle n_{\rm i}/n_{\rm \Lambda CDM}\right\rangle$, removed to emphasise differences in the slope. The residuals at masses above the resolution limit, although noisy, are relatively flat or show little systematic tilt, at least for $z=0$. At $z=0.5$, the coupled cosmology is tilted relative to the uncoupled cosmologies. However, this tilt is not systematic, in one halo, the mass function is steeper, in the other it is flatter. Moreover, this difference is it present at even higher redshifts (not shown here for brevity). Clearly the shape of the subhalo mass distribution, which would be an ideal signature, is primarily governed by nonlinear dynamics (mass loss, dynamical friction, etc.), rather than cosmologically dependent quantities such as the accretion rate or the small difference in the gravitational force experienced by dark matter.

\par 
The top subpanels of this figure also indicate that the differences in the number of and mass fractions within subhaloes seen in \Tableref{tab:bulkproperties} are not informative as the differences in the amplitude of the mass distribution curves are negligible and not correlated with the underlying cosmological model. Coupled cosmologies, at least for $\beta\leq0.05$ are not intrinsically richer or poorer than their uncoupled counterparts.

\subsection{Concentration}
\begin{figure}
    \centering
    \includegraphics[width=0.45\textwidth, trim=1.75cm 20.2cm 11.5cm 2.5cm, clip=true]{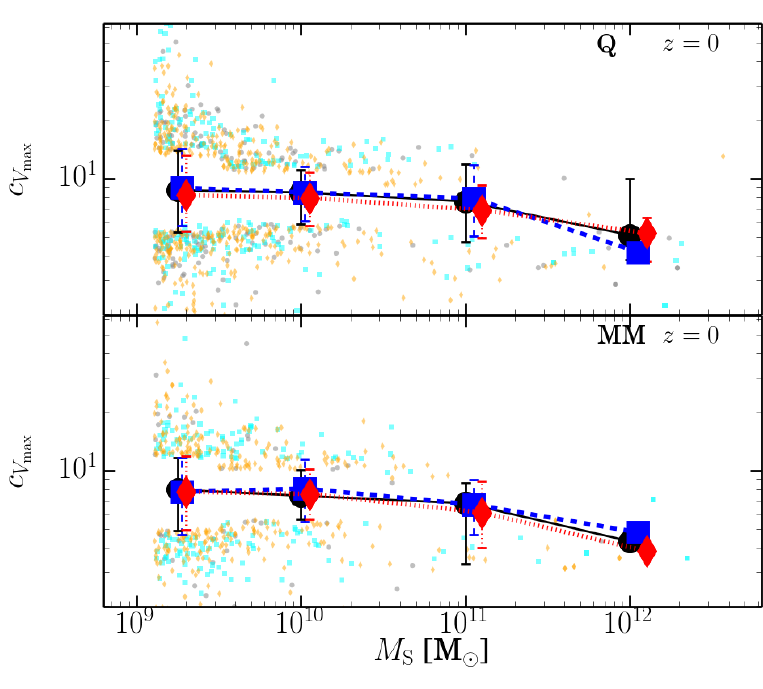}
    \caption{The mass-concentration relation at $z=0$. We first bin the data in mass and calculate the median and the $16\%$ \& $84\%$ quantiles of $c_{\vmax}$ in each bin. These are plotted in the large solid points with error bars. We also plot the outlying points as small points. Line styles and colours are the same as in \Figref{fig:totdenprof:Q}, markers are: \LCDM\ circle, \qCDM\ square, \dedm{0.05} diamond. Outlying point colours are: \LCDM\ grey, \qCDM\ cyan, \dedm{0.05} orange. The x position of the bins are offset by an small arbitrary amount between each cosmology for clarity. As quantiles are misleading for a small sample, we do not plot error bars for bins containing fewer than 5 points, instead, we plot all the points in the bin and the median of these points.}
    \label{fig:subconcen}
\end{figure}

We next examine the concentration of subhaloes in \Figref{fig:subconcen}, specifically the mass dependence of ratio between the virial radius relative to the maximum circular velocity radius, $c_{\vmax}$ (we only show $z=0$ for brevity as $z=0.5$ is similar). There does appear to be a trend for coupled cosmologies to have less concentrated subhaloes, particularly at smaller masses, in agreement with previous studies \cite[][]{baldi2010,carlesi2014a}. The variation in $c_{\vmax}$ at a given mass is large, possibly masking this difference. Moreover, although some of the reduced concentration in the coupled simulation is due to the extra fictional force experienced by dark matter, some of it will also be driven by different expansion histories. The concentration of a halo depends in part on the formation time of the object \cite[e.g.][]{bullock2001,ludlow2014} and is also affected by the dynamical state of a (sub)halo. 

\par
\begin{figure}
    \centering
    \includegraphics[width=0.45\textwidth, trim=1.75cm 20.2cm 11.5cm 2.5cm, clip=true]{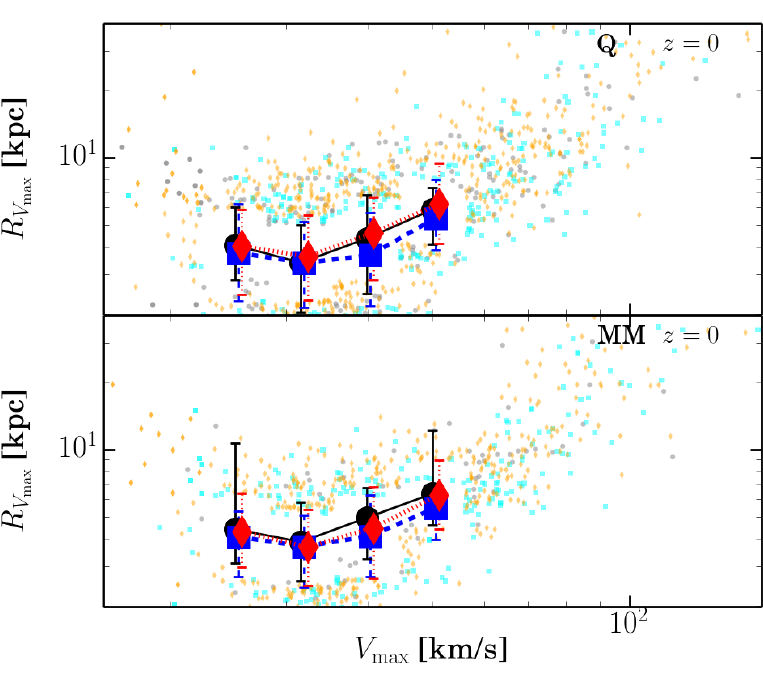}
    \caption{The $\vmax-\rvmax$ distribution similar to \Figref{fig:subconcen}.}
    \label{fig:subvmaxrvmax}
\end{figure}

 
If we turn our attention more directly observable quantities, that is $\vmax-\rvmax$ in \Figref{fig:subvmaxrvmax}, the differences are less evident. Subhaloes are more extended at a given $\vmax$ in \LCDM\ relative to \qCDM. Although including coupling has moved the subhalo distribution closer to the \LCDM\ one, the fact that the quintessence model can differ from the \LCDM\ model suggests that the concentration of subhaloes is not an ideal indicator of a coupled dark sector. 

\subsection{Angular momentum}
\begin{figure}
    \includegraphics[width=0.45\textwidth, trim=1.75cm 17.7cm 11.5cm 2.5cm, clip=true]{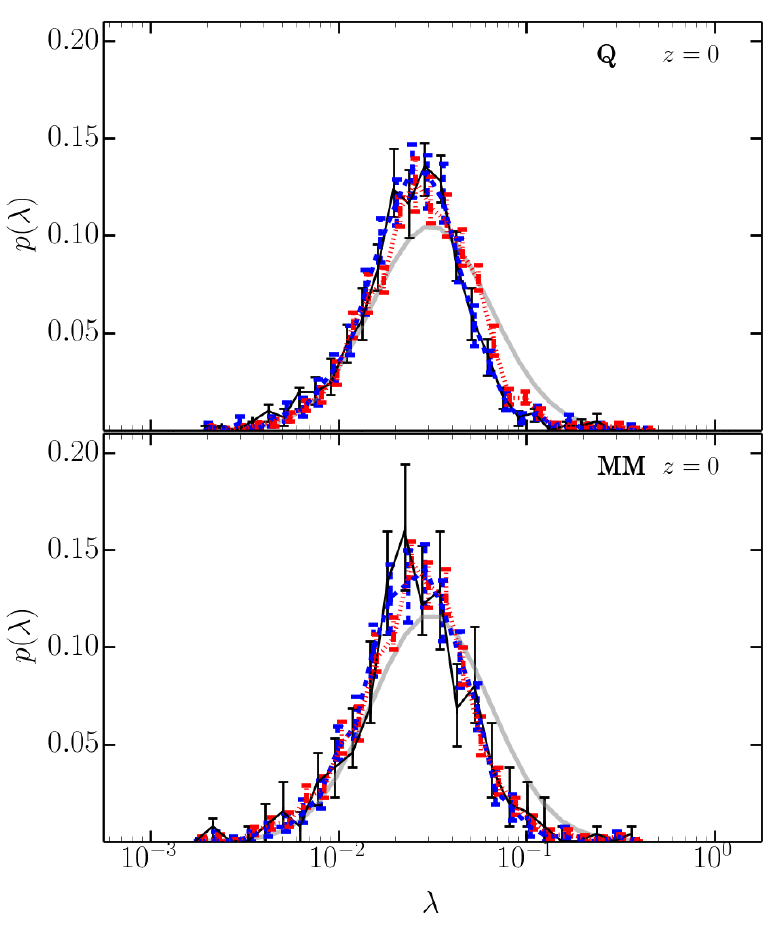}
    \\\vspace{-20.0pt}
    \caption{Subhalo spin distribution at $z=0$. For reference we also show the {\em subhalo} spin distribution from Onions et al (2013) for subhaloes from a dark matter only simulation of a $10^{12}\Msun$ halo by a solid gray line. Onions et al (2013) found that subhaloes peaks at a lower spin value than the {\em halo} distribution. Probabilities and errors are estimate using bootstrap resampling. Line styles and colours are the same as in \Figref{fig:totdenprof:Q}.}
    \label{fig:subspin}
\end{figure}
\begin{figure}
    \centering
    \includegraphics[width=0.45\textwidth, trim=1.75cm 20.2cm 11.5cm 2.5cm, clip=true]{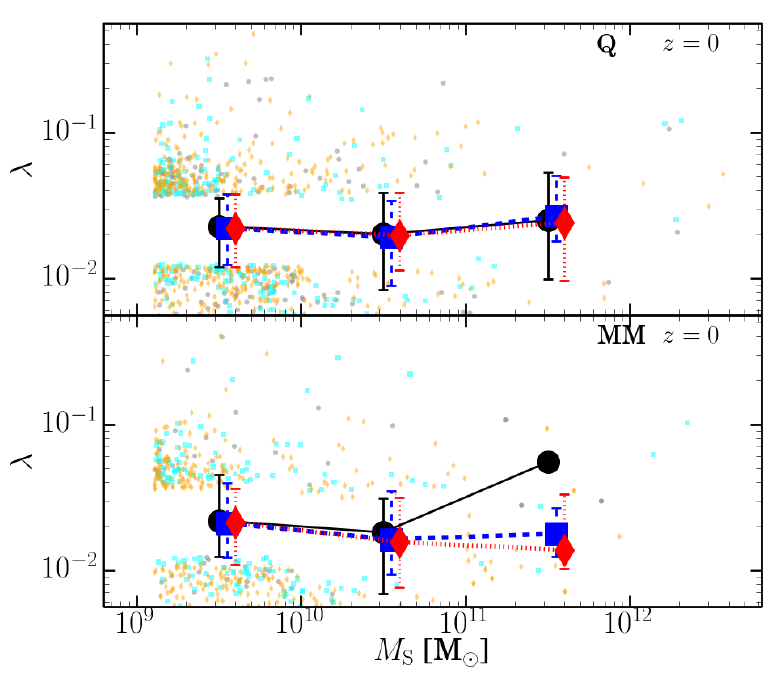}
    \caption{Spin as a function of mass. Format similar to \Figref{fig:subconcen}.}
    \label{fig:subspinmass}
\end{figure}
The angular momentum of field haloes in non-standard cosmologies shows indications that there maybe subtle differences in the peak of the spin parameter distribution \cite[e.g.][]{hellwing2013b,carlesi2014b}. Here we examine the spin parameter as defined by \cite{bullock2001},
\begin{align}
    \lambda=\frac{L}{\sqrt{2}MVR},
\end{align}
where $L$ is the angular momentum, $M$ is the total mass, and $R$ is the radius of the subhalo, and $V^2=GM/R$. Normally, these quantities are measured at the so-called virial radius, $R_\Delta$, however, subhaloes can be truncated at overdensities higher that $\Delta$. For those subhaloes, we use the total mass, angular momentum and size of the object to calculate $\lambda$. 

\par 
Figure \ref{fig:subspin} shows the total distribution across all mass scales. Note that we have calculated spins using the self-bound portions of a substructures, that is we do not include the loosely unbound tidal tails associated with subhaloes and ignore tidal debris, which would produce a longer high spin tail. First we note that the subhalo spin distribution here is lower than that found in \cite{onions2013a}, who studied subhaloes in a galaxy mass host. The spin distribution at higher redshifts and lower host masses is in better agreement with \cite{onions2013a}. 

\par
The major feature of this plot is the absence of systematic difference in the distributions between uncoupled and coupled cosmologies. There is a hint in a shift towards higher spins in this coupled cosmology, there being a small deficit in probability relative to the uncoupled cosmologies just below the peak of the distribution and a possible excess above. However, the differences are not incredibly significant. A simple two-sided Kolmogorov–Smirnov (KS) test comparing the distributions returns p-values of $\gtrsim0.5$, regardless of which two cosmologies are compared, that is the empirical distributions arise from the same parent distribution. This is also true at higher redshifts. Moreover, the difference is more noticeable in Q than in MM, showing the dynamical state of the host does influence the spins of subhaloes residing in it. 

\par
Even separating the distribution into masses, as we do in \Figref{fig:subspinmass}, does not necessarily show an unambiguous signature of a coupled dark sector. Small coupled cosmology subhaloes of have {\em marginally} higher spins. However, even the uncoupled cosmology differs from \LCDM and for the well sampled mass scales, the mass dependence of $\lambda$ shows little difference between cosmologies with the two-sided KS-test giving p-values of $\gtrsim0.4$ regardless of mass bin used. Moreover, the spin parameter is influenced by galaxy formation physics, which we have not included in our simulations. The effect of forming stars is likely to wash out the small differences noted here. 

\subsection{Alignment}
\begin{figure}
    \centering
    \includegraphics[width=0.45\textwidth, trim=1.75cm 18.8cm 11.5cm 2.8cm, clip=true]{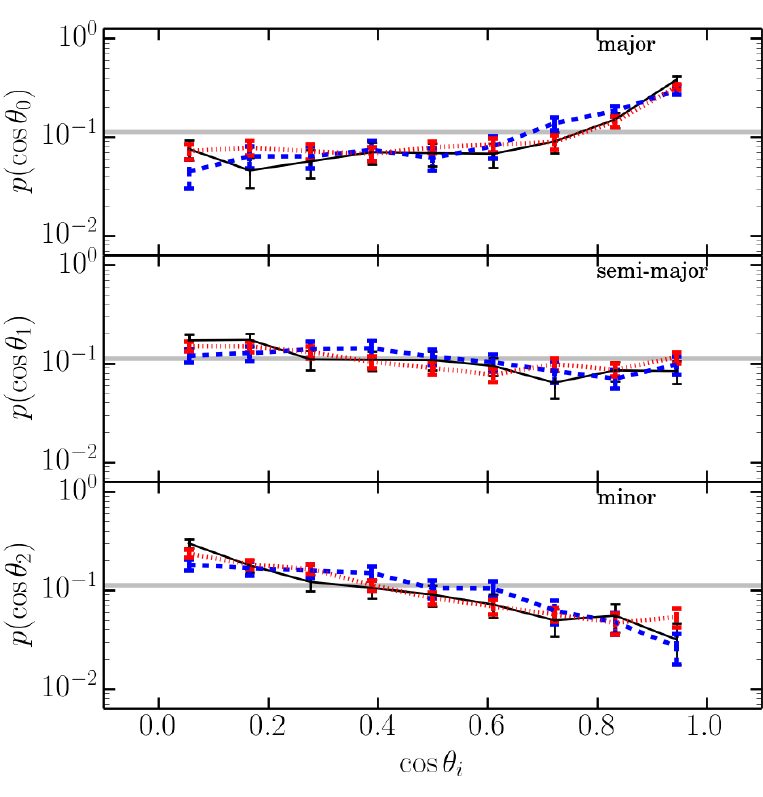}
    \caption{Subhalo alignment with the axes of the host halo for Q at $z=0$. We show the probability for a isotropically distributed distribution by a gray line. Probabilities and errors are estimated using bootstrap resampling. Line styles and colours are the same as in \Figref{fig:totdenprof:Q}.}
    \label{fig:subalign}
\end{figure}
Observations suggests that subhaloes/satellite galaxies are anisotropically distributed within there host halo. \LCDM\ simulations naturally give rise to anisotropic subhalo populations, however, this cosmology struggles to explain the strength of the alignments observed. Here we examine if a coupled dark sector produces a more anisotropically distributed subhalo population via the angle between a subhalo's position, ${\bf x}$, and major, semi-major, and minor axes of the host halo defined by the eigenvectors $e_i$ of the reduced inertia tensor \cite[][]{dubinski1991,allgood2006},
\begin{equation}
    \tilde{I}_{j,k}=\sum\limits_n \frac{m_n x^\prime_{j,n} x^\prime_{k,n}}{(r^\prime_{n})^2}.
    \label{eqn:inertia}
\end{equation}
Here the sum is over particles in the halo, $(r^\prime_n)^2=(x^\prime_n)^2+(y^\prime_n/q)^2+(z^\prime_n/s)^2$ is the ellipsoidal distance between the subhalo's centre of mass and the $n$th particle, primed coordinates are in the eigenvector frame of the reduced inertia tensor, and $q$ \& $s$ are the semi-major and minor axis ratios respectively.

\par 
We show in \Figref{fig:subalign} the distribution of $\cos \theta_i= {\bf x}\cdot{\bf e}_{i}/x$ for Q at $z=0$. This figure indicates subhaloes are typically aligned along the major axis of the host halo. However, haloes in coupled cosmologies do not have subhalo populations that are any more or less anisotropic than those in uncoupled cosmologies. Despite the vary different dynamical state of MM between cosmologies and relative to Q, the overall picture is similar. The subtle difference is that subhaloes are not as strongly aligned/anti-aligned to the major/minor axis of the host halo, a result of the violent phase-mixing in major mergers and randomisation of orbits. This type of analysis in itself does not indicate whether or not coupled cosmologies produce planar arrangement of satellites. 

\par
\begin{figure}
    \centering
    \includegraphics[width=0.45\textwidth, trim=1.75cm 18.8cm 11.5cm 2.8cm, clip=true]{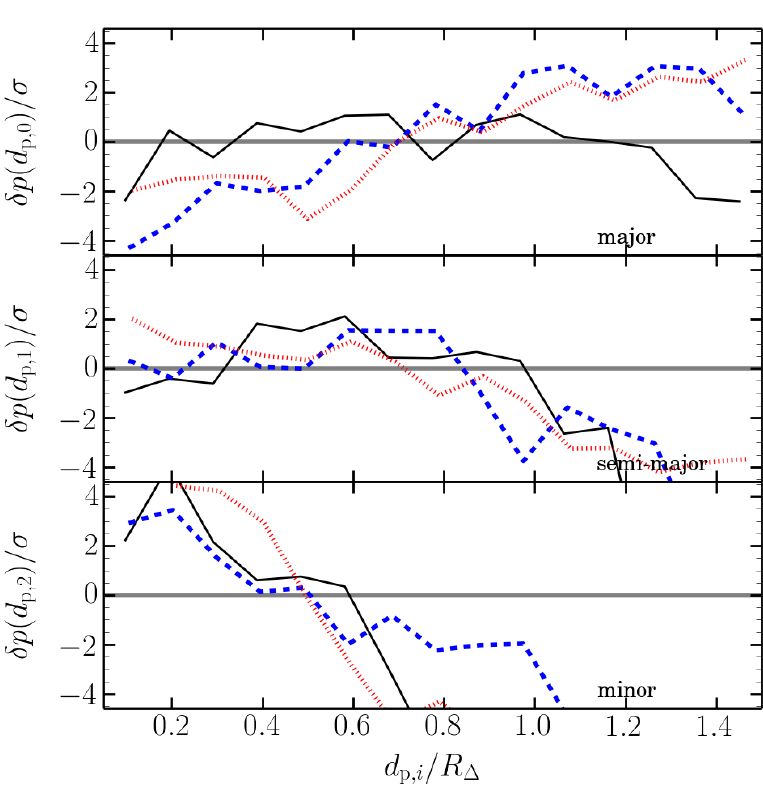}
    \caption{Relative probability distribution of distances to three planes normalised by the variance in the observed PDF for Q at $z=0$. The expectation value for an isotropic distribution is shown by a gray line. Probabilities and errors are estimated using bootstrap resampling. Line styles and colours are the same as in \Figref{fig:totdenprof:Q}.}
    \label{fig:subplane}
\end{figure}

To see if planes are present we examine the distributions of distances to the three planes with normal vectors defined by the inertia tensor's eigenvectors. This distribution is compared to an isotropically distributed subhalo population with the same overall radial distribution present in the halo. We show for Q the ratio relative to the isotropic expectation normalised by the uncertainty, $\delta p/\sigma=(p_{\rm obs}/p_{\rm iso}-1)/\sigma(p_{\rm obs}/p_{\rm iso})$, as function of distance in \Figref{fig:subplane}. 

\par
A planar structure would show up as an excess in the probability of finding subhaloes at small distances from a given plane. Focusing on the ``major'' plane, that which is perpendicular to the major axis that has a normal vector along the major axis, we see that the population is generally consistent with an isotropic distribution (within $2\sigma$). There is a hint of a deficit of objects at small projected distances that is matched by the increase in the number of subhaloes with small projected distances orthogonal to this plane. In general, $\lesssim2\sigma$ deviations occur at all distances from the planes examined here, arising from the overall anisotropic distribution of subhaloes. The exception is the ``minor'' plane, which shows $\gtrsim3\sigma$ deviation away from an isotropic distribution at small projected distances for all three cosmologies. 

\par
Focusing on \LCDM-Q, there are 198 subhaloes ($20\%$ of the subhaloes composed of $\geq100$ particles) within a projected distance of $40$~kpc from the ``minor'' plane spanning a range of radii, from $280$~kpc to $1600$~kpc, that is from $\sim\rvmax$ to past $\sim2R_\Delta$. However, this planar structure is purely spatial, it is not a dynamically stable structure. The angular momentum of the subhaloes not strongly aligned with the normal vector defining the plane ($45^\circ$). Subhalo's are as likely to have tangentially dominated velocities as radial ones and be rotating, counter-rotating or have an orbit principally out of the plane. The velocity dispersions perpendicular to the plane for even the coldest $40\%$ is $\sim200$~km/s, meaning this plane would disperse within $200$~Myr. As expected, \LCDM\ haloes have anisotropic distributions of subhaloes but no planes. 

\par
At first glance the inclusion of coupling looks encouraging. The deviations away from projected distances of an isotropically distributed subhalo population are more significant, although it is still contains similar fractions of the subhalo population as \LCDM ($\approx20\%$). This planar structure has a vertical distribution of $\sim70~$kpc, an angular extent of $\sim180^\circ$, and spans $\sim2~$Mpc in radius. However, its dynamics again show the transient nature of this plane. Its angular momentum axis is misaligned with the plane by $70^\circ$ and the vertical velocities are offset from the plane by $130$~km/s and will disperse within $\sim190$~Myr. The main component of this plane is two groups of subhaloes, which at $z=0.5$ reside in the two main filaments that connect this halo's progenitor to the cosmic web.

\par 
MM, despite being at various points in a merger in the different cosmologies has similar features: anisotropic subhalo distribution and no dynamically stable planar structures. This trend is true for both haloes at earlier redshifts as well. It is however interesting to note that the statistical significance of a flatten spatial distribution in the ``minor'' plane is always higher in the coupled cosmology. 

\par 
That is not to say spatially compact planes do not exist in other cosmologies. We note that at $z=1$ both \qCDM\ \& \dedm{0.05} Q haloes (with $M\approx3.5\times10^{13}\Msun$) have ``minor'' planes of satellites spanning $40$~kpc with an orbital angular momentum axis aligned with the plane's normal vector, although the motions orthogonal to the plane are large enough ($200$~km/s) to disperse the plane within a $\sim200$~Myr. Despite the high redshift, these host masses are closer to the mass scale of the Local Group, where planes have been observed. 

\section{Field Haloes: Coupled cosmologies in lower density environments}
\label{sec:fieldhaloes}
The dense environment within a group appears to show little differences in the subhalo population. However it is worthwhile examining whether this is true in all environments. We therefore examine the bulk properties of haloes that lie outside the group as a function of the distance from the centre of the host. As the haloes are triaxial and we are interested in exploring environments at different densities, we use the ellipsoidal distance to the halo centre, $r^\prime=(x^2+y^2/q^2+z^2/s^2)$, where coordinates are in the eigenvector frame of the reduced inertia tensor, and $q$ \& $s$ are the axis ratios (which for Q are $q\sim0.75$ \& $s\sim0.47$), instead of the pure radial distance.
\begin{figure}
    \centering
    \includegraphics[width=0.45\textwidth, trim=1.75cm 20.2cm 11.5cm 2.5cm, clip=true]{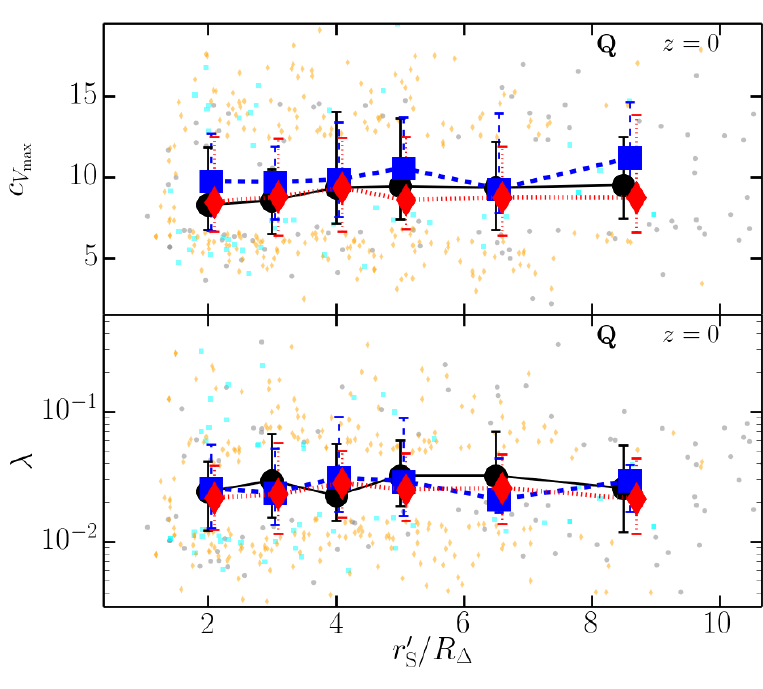}
    \caption{Properties of haloes around Q verses ellipsoidal distance, $r^\prime$. Here we show the spin, $\lambda$ and concentration $c_{\vmax}$ (top and bottom respectively). We select haloes with masses $5\times10^9/\Msun\leq M \leq5\times10^{10}\Msun$, as these quantities depend on host halo mass. Format is similar to \Figref{fig:subconcen}.}
    \label{fig:haloesprop}
\end{figure}

\par 
Figure \ref{fig:haloesprop} shows the concentration and spin of small haloes around Q, the halo with the quiescent merger history. This figure shows haloes differ between cosmologies. Focusing on the top panel of \Figref{fig:haloesprop}, we find haloes are more concentrated in the quintessence cosmology than in \LCDM. Adding coupling has reversed the trend in $c_{\vmax}$, resulting in increasingly less concentrated haloes. The spin distribution to show little dependence on on its environment but this is due to the strict removal of unbound particles. If we include loosely unbound particles within the FOF envelope of field haloes we would clearly see that haloes closer in have higher $\lambda$, a result of becoming tidally disrupted. More importantly, there is little difference between cosmologies in the different environments. These trends apply to MM, despite its active merger history. 

\par
\begin{figure}
    \centering
    \includegraphics[width=0.45\textwidth, trim=1.75cm 20.2cm 11.5cm 2.5cm, clip=true]{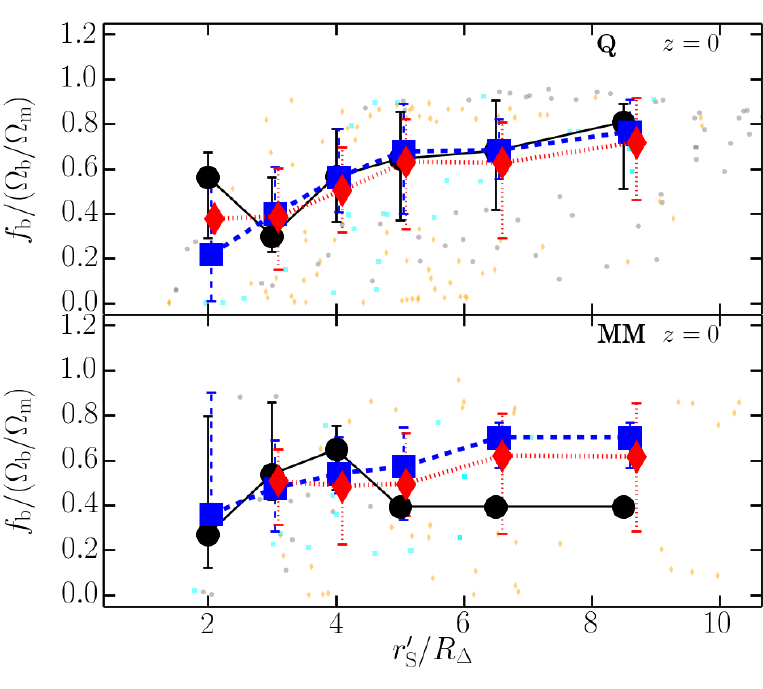}
    \caption{Baryon fraction normalised by the cosmic baryon fraction, $f_{\rm b}/(\Omega_b/\Omega_m)$, for Q (top) \& MM (bottom). Similar to \Figref{fig:haloesprop}.}
    \label{fig:haloesfb}
\end{figure}

As baryons and dark matter do not feel the same gravitational force, we examine the baryon fractions of haloes and the dependence upon environment in \Figref{fig:haloesfb} for both Q \& MM. The first notable feature is the difference between the median $f_{\rm b}$ in Q and MM for the same cosmology. Here the dynamical state and the density of the local environment has affected the median $f_{\rm b}$ at a given $r^\prime$, although the overall trends remain unchanged. Small haloes have decreasing $f_{\rm b}<\Omega_b/\Omega_m$ with decreasing radius as the hot gas surrounding the cluster strips haloes with increasing efficiency the closer a halo is to group centre. Of greater importance is the apparent decrease in $f_{\rm b}$ when $\beta>0$. The exact median baryon fraction depends on the dynamical state but coupling decreases the radius at which there are significant drop in $f_{\rm b}$. The complication lies in the radius at which this occurs. Q haloes show a more gradual change and baryon poor subhaloes appear at much larger distances than in MM. Additionally, the surrounding halo population in \LCDM-MM is strongly effected the the presence of another infalling group mass halo, hence the sudden drop in the median $f_{\rm b}$ and number of subhaloes containing any baryons at all.

\section{Discussion \& Conclusion}\label{sec:discussion} 
We have studied the internal properties of group mass haloes in coupled Dark Energy-Dark Matter cosmologies using adiabatic zoom simulations. The cosmologies used in this study consist of a fiducial \LCDM\ model, a quintessence model, and a strongly coupled DEDM model, \dedm{0.05}. In coupled models, dark matter decays into dark energy, giving rise to the late time accelerated expansion, an additional frictional force experienced by DM, and expansion histories and growth functions that differ from \LCDM. It is important to note that the best fit parameters for a given alternative cosmology to the CMB will differ from the best fit parameters of \LCDM. Given this freedom, we have chosen to pin our simulations at $z=0$ so that the density parameters and $\sigma_8$ match the \LCDM\ values based on \LCDM\ results from Planck. 

\par 
Our approach was to examine in detail two well resolved cluster mass haloes with different dynamical states to identify unambiguous signatures of a coupled dark sector. We find that, considering the differences between \LCDM\ \& \qCDM, the two uncoupled cosmologies, haloes that form in the coupled cosmologies appear remarkably similar to \LCDM\ ones, even in instances where the total mass differs by $80\%$. The density profiles of our two well resolved haloes are well characterised by an NFW (or Einasto) like profile when examined at times when the haloes are relaxed. The baryonic fraction profiles of these two haloes vary significantly but not in a systematic way between cosmologies. 

\par 
The surprising result is the similarity in the subhalo populations of these haloes. The form of the cumulative subhalo mass function is seemingly unaffected by the inclusion of extra dark sector physics. We do find that both subhaloes and small haloes in the outskirts of the group are less concentrated on average in coupled cosmologies. If we look at differences in the $\vmax-\rvmax$ plane, which is more directly observable, we find subhaloes with small circular velocities are typically more extended in coupled cosmologies. Unfortunately, a halo's concentration partially depends on its formation time and thus depends on the choice of cosmological parameters and when a comparison between cosmologies is made. Furthermore, as the concentration of subhaloes can drastically change with the inclusion of galaxy formation physics, which we did not include, this probe of a coupled dark sector is not ideal.

\par 
We have also examined these haloes to see if a spatially compact planar arrangement of satellites is present given the observed distributions of satellites around the only two galaxies in which these types of observations are possible, our Galaxy and M31 \cite[e.g.][]{pawlowski2012,ibata2013,conn2013}. In all the cosmologies studied here, we do not find strong evidence for a subpopulation of satellites residing in a stable co-rotating plane. However, we do find for both haloes at multiple epochs, our coupled cosmology is more likely to have a statistically significant ``planes''. Unfortunately, these are not dynamically stable and will dispersing in roughly $\sim200$~Myr. These dynamically unstable planar structures are a natural outcome of cosmic structure formation \cite[e.g.][]{sawala2015a,libeskind2015a}: Subhaloes are principally accreted along filaments. The current tension with observations lies in the how compact these planes are (thicker than observations suggest) and the fact that they do not have strongly correlated velocities. However, the increase in the statistical significance of planar like arrangement of subhaloes in strongly coupled groups is intriguing and warrants further study. Higher filamentary accretion should increase the likelihood of observing planar structures. Thus clues for a coupled dark sector may lie in the amount of material accreted along filaments. Whether there is an increase the observed velocity correlation between satellites relative to \LCDM\ and in better agreement with observation of \cite{ibata2014b} remains to be seen. A complete study of the growth of filaments and the accretion history of haloes is beyond the scope of this paper and will be pursued in the future. 

\par
We have also looked for signatures of a coupled dark sector in haloes residing well outside the highly nonlinear cluster environment. Although these haloes show the same trends in their bulk properties as a function of cosmology, i.e., lower concentrations in coupled cosmologies, there is little dependence on environment unique to coupled dark sector physics. The only environmental signature noted lies in the baryon fractions: outlying haloes of a cluster are more likely to have lower $f_{\rm b}$ than their uncoupled counterparts at the same ellipsoidal distance or same environment. The distance (or density) at which there is a significant decrease in the average baryon fractions occurs at moderately larger radii in coupled cosmologies. However, this signature is subtle and is not noticeable when examining baryon fractions as a function of radial (or projected distance) distance to the group centre. Moreover, we have neglected all the complexity associated with galaxy formation physics and the exact treatment of gas and subgrid physics significantly effects the ability of similar mass haloes to retain gas in a given environment \cite[e.g.][]{tasker2008,scannapieco2012,nifty1}.

\par
Thus, the internal properties of haloes are seemingly indifferent to a coupled dark sector, presenting an interesting possibility: that moderately coupled dark sector models are viable and this physics could just be hidden from view.

\section*{Acknowledgements}
The authors would like to thank the referee for a quick reply and their comments. PJE is supported by the SSimPL programme and the Sydney Institute for Astronomy (SIfA), DP130100117 and DP140100198. CP is supported by DP130100117, DP140100198, and FT130100041. GFL acknowledges financial support through DP130100117. AK is supported by the {\it Ministerio de Econom\'ia y Competitividad} (MINECO) in Spain through grant AYA2012-31101 as well as the Consolider-Ingenio 2010 Programme of the {\it Spanish Ministerio de Ciencia e Innovaci\'on} (MICINN) under grant MultiDark CSD2009-00064. He also acknowledges support from the {\it Australian Research Council} (ARC) grants DP130100117 and DP140100198. He further thanks Combustible Edison for shizophonic. This research was undertaken on the NCI National Facility in Canberra, Australia, which is supported by the Australian commonwealth Government and with resources provided by Intersect Australia Ltd.

\pdfbookmark[1]{References}{sec:ref}
\bibliographystyle{mn2e}
\bibliography{dedmsubstructure.bbl}

\end{document}